\documentclass[12pt]{article}
\include{epsf}
\include{psfig}
\include{epsfig}
\usepackage{epsfig}
\makeatletter
\def\gsim{\compoundrel>\over\sim}
\def\lsim{\compoundrel<\over\sim}
\def\compoundrel#1\over#2{\mathpalette\compoundreL{{#1}\over{#2}}}
\def\compoundreL#1#2{\compoundREL#1#2}
\def\compoundREL#1#2\over#3{\mathrel
         {\vcenter{\hbox{$\m@th\buildrel{#1#2}\over{#1#3}$}}}}
\makeatother

\makeatletter
\begin{document}
\setcounter{page}{0}
\thispagestyle{empty}
\setlength{\parindent}{1.0em}
\begin{flushright}
GUTPA/03/07/01
\end{flushright}
\renewcommand{\thefootnote}{\fnsymbol{footnote}}
\begin{center}{\LARGE{{\bf  The Origin of Mass }}}\end{center}
\begin{center}{\large{C. D.
Froggatt}~\footnote[2]{E-mail: c.froggatt@physics.gla.ac.uk}\\}
\end{center}
\renewcommand{\thefootnote}{\arabic{footnote}}
\begin{center}{{\it Department of Physics and Astronomy}\\{\it
University of Glasgow, Glasgow G12 8QQ, Scotland}}\end{center}
\setcounter{footnote}{0}

\begin{abstract}
The quark-lepton mass problem and the ideas of mass protection are
reviewed. The hierarchy problem and suggestions for its
resolution, including Little Higgs models, are discussed. The
Multiple Point Principle is introduced and used within the
Standard Model to predict the top quark and Higgs particle masses.
Mass matrix ans\"{a}tze are considered; in particular we discuss
the lightest family mass generation model, in which all the quark
mixing angles are successfully expressed in terms of simple
expressions involving quark mass ratios. It is argued that an
underlying chiral flavour symmetry is responsible for the
hierarchical texture of the fermion mass matrices. The
phenomenology of neutrino mass matrices is briefly discussed.
\end{abstract}

\vspace {3cm}

{\it To be published in the Proceedings of the XXXI ITEP Winter 
School of Physics, Moscow, Russia, 18 - 26 February 2003.}
\newpage
%%%%%%%%%%%%%%%%%%%%%%%%%%%%%%%%%%%%%%%%%%%%%%%%%%%%%%%

\section{Introduction}
\label{introduction}

The origin of the quark and lepton masses, their mixing and three
generation structure remains the major outstanding problem in
particle physics. The charged fermion masses and mixing angles
derive from Yukawa couplings, which are arbitrary parameters
within the Standard Model (SM). Furthermore the non-vanishing
neutrino masses and mixings are direct evidence for physics beyond
the SM. So the experimental values of the fermion masses and
mixings provide our best clue to this new physics. The main
features requiring explanation are the following:
\begin{enumerate}
\item The large mass ratios between generations: \newline
$m_u \ll m_c \ll m_t ; \quad m_d \ll m_s \ll m_b ; \quad m_e \ll
m_{\mu} \ll m_{\tau}$.
\item The large mass splitting within the third (heaviest)
generation: \newline $m_{\tau} \sim m_b \ll m_t$.
\item The smallness of the off-diagonal elements of
the quark weak coupling matrix $V_{CKM}$.
\item The tiny neutrino masses and the large off-diagonal elements
of the lepton weak coupling matrix $U_{MNS}$.
\end{enumerate}

The charged fermion mass hierarchy ranges over five orders of
magnitude, from $\frac{1}{2}$ MeV for the electron to 174 GeV for
the top quark. It is only the top quark which has a mass of order
the electroweak scale $\langle\phi_{WS}\rangle$ = 246 GeV and a
Yukawa coupling of order unity. All of the other charged fermion
masses are suppressed relative to the natural scale of the SM. The
main problem in understanding the charged fermion spectrum is not
why the top quark is so heavy but rather why the electron is so
light. Indeed, as the top quark mass is the dominant term in the
fermion mass matrix, it is likely that its value will be
understood dynamically before those of the other fermions. In
contrast there seems to be a relatively mild neutrino mass
hierarchy and two of the leptonic mixing angles are close to
maximal ($\theta_{atmospheric} \simeq \pi/4$ and $\theta_{solar}
\simeq \pi/6$). The absolute neutrino mass scale ($m_{\nu} < 1$
eV) is another puzzle, suggesting a new physics mass scale -- the
so-called see-saw scale $\Lambda_{seesaw} \sim 10^{15}$ GeV.

We give an overview of the quark-lepton spectrum in section
\ref{spectrum} and introduce the mechanism of mass protection by
approximately conserved chiral charges in section \ref{protection}.
Various approaches to the gauge hierarchy problem are considered
in section \ref{hierarchyproblem}, including supersymmetry, Little
Higgs models and the Multiple Point Principle. In section
\ref{top} we discuss the connection between the top quark and
Higgs masses and how they might be determined dynamically.
Ans\"{a}tze for the texture of fermion mass matrices and the
resulting relationships between masses and mixing angles are
considered in section \ref{ansatze}. Mass protection is proposed
as a natural explanation for the origin of fermion mass matrix
texture in section \ref{origin}. The neutrino mass problem is
discussed in section \ref{neutrino} and finally, in section
\ref{conclusion}, we present a brief conclusion.

\section{The Quark-Lepton Spectrum}
\label{spectrum}

The physical masses of the charged leptons can be directly
measured and correspond to the poles in their propagators:
\begin{equation}
M_e = 0.511 \ \makebox{MeV} \qquad M_{\mu} = 106 \ \makebox{MeV}
\qquad M_{\tau} = 1.78 \ \makebox{GeV}
\end{equation}
However, due to confinement, the quark masses cannot be directly
measured and have to be extracted from the properties of hadrons.
Various techniques are used, such as chiral perturbation theory,
QCD sum rules and lattice gauge theory. The quark mass parameters
extracted from the data usually depend on a renormalisation scale
$\mu$ and the corresponding running masses $m_q(\mu)$ are related
to the propagator pole masses $M_q$ by
\begin{equation}
M_q = m_q(\mu=m_q)\left[ 1+ \frac{4}{3}\alpha_3(m_q) \right]
\end{equation}
to leading order in QCD. The light $u$, $d$ and $s$ quark masses
are usually normalised to the scale $\mu=1$ GeV, corresponding to
the non-perturbative scale of dynamical chiral symmetry breaking
(or $\mu=2$ GeV for lattice measurements), and are typically given
\cite{pdg} as follows:
\begin{eqnarray}
m_u(1 \ \makebox{GeV}) & = & 4.5 \pm 1 \ \makebox{MeV} \nonumber \\
m_d(1 \ \makebox{GeV}) & = & 8 \pm 2 \ \makebox{MeV} \nonumber \\
m_s(1 \ \makebox{GeV})& = & 150 \pm 50 \ \makebox{MeV}
\end{eqnarray}
However the renormalisation scale for the heavy quark masses is
taken to be the quark mass itself which is in the perturbative
regime:
\begin{eqnarray}
m_c(m_c) & = & 1.25 \pm 0.15 \ \makebox{GeV} \nonumber \\
m_b(m_b) & = & 4.25 \pm 0.15 \ \makebox{GeV} \nonumber \\
m_t(m_t) & = & 166 \pm 5 \ \makebox{GeV}
\end{eqnarray}
Note that the top quark mass, $M_t = 174 \pm 5$ GeV, measured at
FermiLab is interpreted as the pole mass.

The quark masses, of course, arise from the diagonalisation of the
three generation mass matrices $M_U$ and $M_D$, by performing unitary
transformations\footnote{Note that with an appropriate choice of the
unitary matrices, it is possible to arrange that the mass eigenvalues
$m_i$ are real and positive for arbitrary mass matrices.} on the
left-handed and right-handed quarks respectively:
\begin{eqnarray}
 U_UM_UV_U^{\dagger} & = & \mbox{Diag} (m_u, m_c, m_t)  \nonumber \\
 U_DM_DV_D^{\dagger} & =  & \mbox{Diag} (m_d,m_s,m_b)
\end{eqnarray}
The quark mixing matrix
\begin{equation}
V_{CKM} = U_UU_D^{\dagger}
\end{equation}
is a measure of the difference between the unitary transformations
$U_U$ and $U_D$ acting on the left-handed up-type and down-type
quarks and has also been measured:
\begin{equation}
 |V_{CKM}| = \pmatrix
{0.9734 \pm 0.0008 & 0.2196 \pm 0.0020 & 0.0036 \pm 0.0007 \cr
0.224 \pm 0.016   & 0.996 \pm 0.013   & 0.0412 \pm 0.002 \cr
0.0077 \pm 0.0014 & 0.0397 \pm 0.0033 & 0.9992 \pm 0.0002 \cr}
\end{equation}
Due to the arbitrariness in the phases of the quark fields, the
mixing matrix $V_{CKM}$ contains only one CP violating phase
\cite{vysotsky},which is of order unity:
\begin{equation}
\sin^2\delta_{CP} \sim 1
\end{equation}

From solar and atmospheric neutrino oscillation data
\cite{simpson}, we know the neutrino mass squared differences:
\begin{equation}
 \Delta m_{21}^2 \sim 5 \times 10^{-5} \qquad
 \Delta m_{32}^2 \sim 3 \times 10^{-3}
 \label{dm2}
\end{equation}
However we do not know the absolute neutrino masses although there
is a similar upper limit of $m_{\nu_i} \lsim 1$ eV from tritium beta decay
and from cosmology. Neutrino
oscillation data also constrain the magnitudes of the lepton
mixing matrix elements to lie in the following $3\sigma$ ranges
\cite{garcia}:
\begin{equation}
 |U_{MNS}| = \pmatrix
 {0.73-0.89 & 0.45-0.66 & <0.24     \cr
 0.23-0.66  & 0.24-0.75 & 0.52-0.87 \cr
 0.06-0.57  & 0.40-0.82 & 0.48-0.85 \cr}
 \label{mns}
\end{equation}
Due to the Majorana nature of the neutrino mass matrix, there are
three unknown CP violating phases $\delta$, $\alpha_1$ and
$\alpha_2$ in this case \cite{kayser}. Note that, unlike the quark
mixing matrix, $U_{MNS}$ is {\em not} hierarchical; all its
elements are of the same order of magnitude except for
$|U_{e3}|<0.24$.

\section{Fermion Mass and Mass Protection}
\label{protection}

A fermion mass term
\begin{equation}
{\cal{L}}_{mass} =  -m \overline{\psi}_L \psi_R + h. c.
\end{equation}
couples together a left-handed Weyl field $\psi_L$ and a
right-handed Weyl field $\psi_R$, which then satisfy the Dirac
equation
\begin{equation}
 i\gamma^{\mu} \partial_{\mu} \psi_L = m \psi_R
\end{equation}
If the two Weyl fields are not charge conjugates $\psi_L \neq
(\psi_R)^c$ we have a Dirac mass term and the two fields $\psi_L$
and $\psi_R$ together correspond to a Dirac spinor. However if the
two Weyl fields are charge conjugates $\psi_L = (\psi_R)^c$ we
have a Majorana mass term and the corresponding four component
Majorana spinor has only two degrees of freedom. Particles
carrying an exactly conserved charge, like the electron carries
electric charge, must be distinct from their anti-particles and
can only have Dirac masses with $\psi_L$ and $\psi_R$ having equal
charges. However a neutrino could be a massive Majorana particle.

If $\psi_L$ and $\psi_R$ have different quantum numbers,
i.e.~belong to inequivalent representations of a symmetry group
$G$ ($G$ is then called a chiral symmetry), a Dirac mass term is
forbidden in the limit of an exact $G$ symmetry and they represent
two massless Weyl particles. Thus the $G$ symmetry ``protects'' the
fermion from gaining a mass. Such a fermion can only gain a mass
when $G$ is spontaneously broken.

The left-handed and right-handed top quark, $t_L$ and $t_R$ carry
unequal Standard Model $SU(2) \times U(1)$ gauge charges
$\vec{Q}$:
\begin{equation}
 \vec{Q}_L \neq \vec{Q}_R \qquad \mathrm{(Chiral\  charges)}
\end{equation}
Hence electroweak gauge invariance protects the quarks and leptons
from gaining a fundamental mass term ($\overline{t}_L t_R$ is not
gauge invariant). This {\em mass protection} mechanism is of
course broken by the Higgs effect, when the vacuum expectation
value of the Weinberg-Salam Higgs field
\begin{equation}
 <\phi_{WS}> = \sqrt{2} v = 246 \ GeV
\end{equation}
breaks the gauge symmetry and the SM gauge invariant Yukawa
couplings $y_i$ generate the running quark masses $m_i$, such as:
\begin{eqnarray}
 m_t = y_t \frac{<\phi_{WS}>}{\sqrt{2}} = y_t \, v
 = y_t 174 \ \makebox{GeV} \\
 m_b = y_b \frac{<\phi_{WS}>}{\sqrt{2}} = y_b \, v
 = y_b 174 \ \makebox{GeV}
\end{eqnarray}
for the top and bottom quarks. In this way a top quark mass of the
same order of magnitude as the SM Higgs field vacuum expectation
value (vev) is naturally generated (with $y_t$ unsuppressed). Thus
the Higgs mechanism explains why the top quark mass is suppressed,
relative to the fundamental (Planck, GUT...) mass scale of the
physics beyond the SM, down to the scale of electroweak gauge
symmetry breaking. However the further suppression of the other
quark-lepton masses ($y_b$, $y_c$, $y_s$, $y_u$, $y_d$ $\ll$ 1)
remains a mystery, which it is natural to attribute to mass
protection by another approximately conserved chiral (gauge)
charge (or charges) beyond the SM, as discussed in section
\ref{origin}.

We remark here that, in the Minimal Supersymmetric Standard Model
(MSSM), the SM Higgs field and its complex conjugate are replaced
by two Higgs fields:
\begin{eqnarray}
 \phi_{WS} \rightarrow H_D \qquad v_1 = \frac{<H_D>}{\sqrt{2}}
 = v \cos \beta \nonumber \\
 \phi_{WS}^{\dagger} \rightarrow H_U \qquad v_2 =
 \frac{<H_U>}{\sqrt{2}}
 = v \sin \beta
\end{eqnarray}
So, in the MSSM, the top and bottom quark masses are expressed as
follows in terms of their Yukawa coupling constants:
\begin{eqnarray}
 m_t = y_t \frac{<H_U>}{\sqrt{2}} = y_t v_2 = y_t \, 174 \sin \beta \
 \makebox{GeV} \\
 m_b = y_b \frac{<H_D>}{\sqrt{2}} = y_b v_1 = y_b \, 174 \cos \beta \
 \makebox{GeV}
\end{eqnarray}

Fermions which are vector-like under the SM gauge group
($\vec{Q}_L = \vec{Q}_R$) are not mass protected and are expected
to have a large mass associated with new (grand unified,
string,..) physics. The Higgs particle, being a scalar, is not
mass protected and {\em a priori} would also be expected to have a
large mass; this is the well-known gauge hierarchy problem.

\section{The Hierarchy Problem}
\label{hierarchyproblem}

\begin{figure}[t]
\leavevmode
%\vspace{-4cm}
\centerline{ \epsfxsize=3.75cm \epsfbox{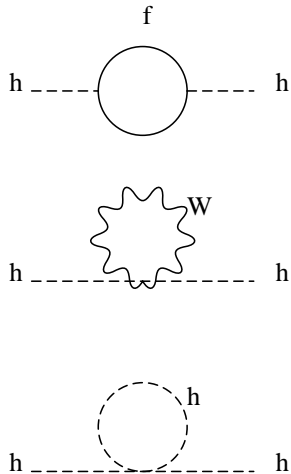} }
\caption{Quadratically divergent contributions to the Higgs mass
in the Standard Model: top loop, electroweak gauge boson loops and
Higgs loop.} \label{fig:jdhiggs}
\end{figure}

As just discussed, there is no symmetry within the SM model which
protects the Higgs particle from acquiring a mass associated with
physics beyond the SM. The Higgs boson mass squared gets
corrections depending quadratically on the SM ultra-violet (UV)
cut-off $\Lambda$ from the one-loop diagrams of
Fig.~\ref{fig:jdhiggs}.
However precision data from LEP indicate that the Higgs mass lies
in the range 114 GeV $< M_h <$ 200 GeV. Therefore the sum of the
bare mass squared term and the radiative corrections $\sim
\Lambda^2$ from Fig.~\ref{fig:jdhiggs} must give a mass in this
range. This leads to a fine-tuning problem for $\Lambda >1$ TeV
when, for example, the magnitude of the top loop contribution
exceeds the physical Higgs mass. We now briefly discuss some
proposed solutions to this problem.

\subsection{Supersymmetry}
\label{susy}

\begin{figure}[t]
\leavevmode
%\vspace{-4cm}
\centerline{ \epsfxsize=10.0cm \epsfbox{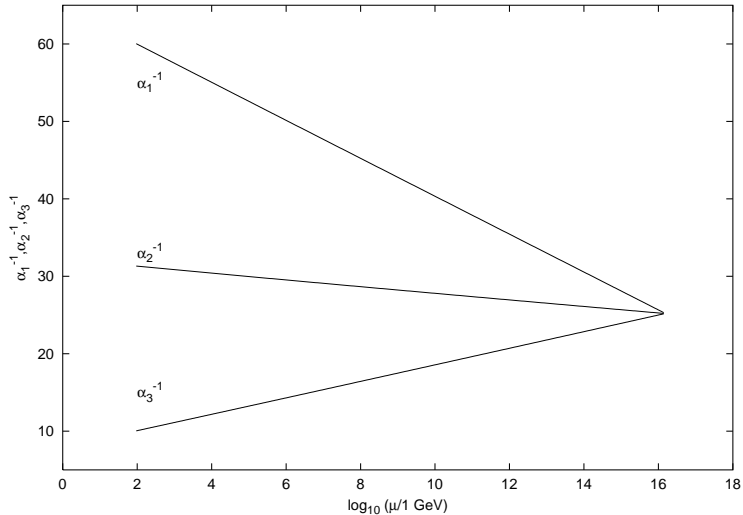} }
\caption{The running of the inverse fine structure constants in
the MSSM.} \label{fig:jdMSSM}
\end{figure}

The most popular approach to solving the hierarchy problem is
based on supersymmetry (SUSY). Indeed the popularity of SUSY is
largely based on its success in solving this problem and the
consistency of the Minimal Supersymmetric Standard Model (MSSM)
with the supersymmetric grand unification of the $SU(3) \times
SU(2) \times U(1)$ running gauge coupling constants at a scale
$\mu = \Lambda_{GUT} \sim 3 \times 10^{16}$ GeV, as illustrated in
Fig.~\ref{fig:jdMSSM}.
In SUSY, essentially all the quadratically divergent loop diagrams
in Fig.~\ref{fig:jdhiggs} have corresponding superpartner
diagrams, involving stop, gaugino and higgsino loops. In the limit
of exact supersymmetry, the diagrams cancel completely reflecting
the opposite sign between fermion and boson loops. However in the
MSSM, it is supposed that SUSY is softly broken at a scale $\mu =
M_{SUSY} \sim 1$ TeV. So, with typical superpartner masses of
order $M_{SUSY}$, the cancellation is incomplete and the SM
cut-off $\Lambda$ is replaced by $M_{SUSY}$.

It follows that SUSY solves the technical hierarchy problem, in
the sense that once the ratio $\frac{M_h^2}{\Lambda_{GUT}^2}$ is
set to be of order $10^{-28}$ in the MSSM at tree level, then it
remains of order $10^{-28}$ to all orders of perturbation theory.
Thus SUSY stabilizes the hierarchy between the electroweak and
grand unified scales, but does not explain why the gauge hierarchy
exists in the first place. This is reflected in the so-called
doublet-triplet splitting problem of grand unified theories
(GUTs). In the simplest GUT, the Weinberg-Salam Higgs doublet
field combines with a coloured triplet Higgs field to form a
5-plet of SU(5). The Higgs doublet remains light of order 100 GeV,
whereas the triplet Higgs contributes to proton decay and must
therefore have a mass of order $10^{16}$ GeV. In the minimal SUSY
SU(5) model, this requires a fine-tuning of parameters with an
accuracy of 1 part in $10^{14}$.

\subsection{Composite Higgs and Pseudo-Goldstones}

The idea of replacing the elementary Higgs scalar field by
composite pseudo-Goldstone bosons is motivated by the success of QCD as
a theory of strong interactions. QCD neatly avoids the hierarchy
problem, by the very definition of the scale of strong
interactions $\Lambda_{QCD} \sim 200$ MeV as the scale at which
the asymptotically free QCD fine structure constant
$\alpha_3(\mu)$ diverges. The SM one loop renormalisation group
equation for $\alpha_3(\mu)$
\begin{equation}
16\pi^2\frac{dg_3}{d \ln \mu} = b_3g_3^2 \qquad b_3 = -7
\end{equation}
generates a logarithmic dependence on the energy
scale $\mu$. Consequently with an input value at the Planck
scale, $\Lambda_{Planck} \sim 10^{19}$ GeV, for the QCD
coupling constant $g_3$ of order unity,
$g_3(\Lambda_{Planck}) =0.7$ corresponding to
$\alpha_3(\Lambda_{Planck})=g_3^2(\Lambda_{Planck})/4\pi \sim .04$, the
QCD scale is naturally generated to be
$\Lambda_{QCD} = \Lambda_{Planck} \exp (-32\pi^2/7) \sim 200$
MeV, where $\alpha_3(\Lambda_{QCD}) \rightarrow \infty$. Thus the
hierarchy $\Lambda_{QCD}/\Lambda_{Planck} \sim 4 \times 10^{-20}$ is
obtained without fine-tuning.

The confining QCD force is responsible for the formation of a
$\overline{q}q$
quark condensate in the vacuum, which spontaneously breaks the
approximate global $G = SU(2)_L \times SU(2)_R$ chiral symmetry
associated with the light $u$ and $d$ quarks ($m_u,m_d \ll
\Lambda_{QCD}$) down to the diagonal isospin group $H =
SU(2)_{L+R}$ of hadronic physics. Thus three composite
pseudo-Goldstone bosons -- the pions -- arise as a result of this
spontaneous breakdown of $\dim G - \dim H = 3$ of the global
symmetry generators; the pions are not exactly massless Goldstone
bosons because the global symmetry is explicitly broken by the
small quark masses $m_u$ and $m_d$. The quark condensate also
spontaneously breaks the electroweak $SU(2) \times U(1)$ gauge
symmetry. So, in the absence of the Higgs
field, the pions would become the longitudinally polarised weak
gauge boson states, generating a mass for say the $W$ boson of
$\frac{1}{2}g_2f_{\pi} \sim 30$ MeV, where $f_{\pi} \sim
\Lambda_{QCD}/2 \sim 95$ MeV is the pion decay constant. However,
in the SM, there is an elementary Higgs field and the pions remain
as physical pseudoscalar mesons, whose low
energy dynamics is well described by chiral perturbation theory
based on the $G/H$ non-linear sigma model. This non-linear sigma
model becomes strongly coupled at an UV cut-off scale $\Lambda
\sim 4\pi f_{\pi} \sim 1$ GeV, where non-Goldstone hadronic bound
states and resonances are formed.

In technicolor models \cite{technicolor}, electroweak symmetry
breaking is provided by an asymptotically free $SU(N_{TC})$ gauge
theory without elementary scalars, analogous to QCD, with a
confinement scale of order the electroweak scale:
\begin{equation}
 \Lambda_{TC} \sim v = 174 \ GeV
\end{equation}
Massless technifermions, analogous to the up and down quarks, form
vacuum condensates spontaneously breaking the electroweak gauge
symmetry and generating massless Goldstone bosons known as
technipions. These technipions provide the longitudinal $W$ and
$Z$ gauge boson states and their physical masses. The generated
$W$ boson mass is given in this case by $M_W = \frac{1}{2} g_2
f_{TC} \sim 80$ GeV, where $f_{TC} = \sqrt{2}v =246$ GeV is the
technipion decay constant. So the hierarchy problem is naturally
solved without fine-tuning. The UV cut-off for the non-linear
sigma model description of the technipion interactions is given by
$\Lambda \sim 4\pi f_{TC} \sim 2$ TeV, which is the scale where
the new technihadron physics enters. However the simple
technicolor mechanism does not communicate the electroweak
symmetry breaking to the quarks and leptons, leaving them
massless. In order to generate quark and lepton masses, it is
necessary to complicate the model significantly, introducing
phenomenological problems with flavour changing neutral currents
and precision electroweak data.

There have also been attempts to identify the Higgs boson as a
bound state of a $t$ and a $\overline{t}$ quark, as a consequence
of a large Yukawa coupling of the top quark and the formation of a
top quark condensate. Top quark condensation models lead to the
infrared fixed point prediction for the top quark mass (see
section \ref{infrared}).

The idea of identifying the Higgs boson itself as a
pseudo-Goldstone boson has recently been revived in the so-called
Little-Higgs models \cite{little}. The new ingredient is to
arrange that the UV cut-off $\Lambda \sim 4\pi f$ of the
associated $G/H$ non-linear sigma model is postponed to 10 TeV
without fine-tuning, corresponding to a pseudo-Goldstone decay
constant $f$ of order 1 TeV. It is hoped that the higher (10 TeV)
scale for the new physics, underlying the $G \rightarrow H$ global
symmetry breaking, is sufficient to avoid it giving
phenomenological problems associated with precision data and
flavour changing neutral currents. The non-zero masses for the
pseudo-Goldstone bosons are generated by the gauge, Yukawa and
scalar couplings, which explicitly break the global symmetries. In
order to obtain a realistic light Higgs scenario, it is necessary
that the one loop quadratically divergent contributions to the
Higgs mass in Fig.~\ref{fig:jdhiggs} are naturally cancelled by
weakly coupled physics at the scale $f$. These cancellations
involve one loop diagrams containing new particles, related to the
top quark, the W and Z gauge bosons and the Higgs particle, having
appropriate coupling constant values. Furthermore, in contrast to
supersymmetry, they occur between particles with the same
statistics. As a consequence of these cancellations, quadratically
divergent corrections to the Higgs mass arise only at two and
higher loop order, making the small Higgs mass natural and
motivating the name ``Little Higgs''. This Little Higgs scenario
requires that the gauge, Yukawa and scalar couplings are organised
in such a way that each one separately preserves enough of the
global symmetry to forbid the Higgs mass. Then two or more
different couplings are required to act together, in two or higher
loop diagrams, in order to generate quadratically divergent
contributions to the Higgs mass.

As a concrete example, we outline the structure of the simplest
model based on an $SU(5)/SO(5)$ non-linear sigma model and known
as the Littlest Higgs model \cite{littlest}. The $SU(5)$ global symmetry
contains the extended gauge group $\left[ SU(2) \times U(1)
\right]^2$. At the same time as the global symmetry breakdown,
$SU(5) \rightarrow SO(5)$, the gauge symmetry $\left[ SU(2) \times
U(1) \right]^2$ breaks down to its diagonal subgroup $SU(2) \times
U(1)$ identified as the SM electroweak gauge group. The global
symmetry breakdown results in 14 Goldstone bosons, which decompose
under the electroweak gauge group $SU(2) \times U(1)$ as a real
singlet $\mathbf{1}_0$, a real triplet $\mathbf{3}_0$, a complex
doublet $\mathbf{2}_{\pm 1/2}$ and a complex triplet
$\mathbf{3}_{\pm 1}$. The real singlet and triplet bosons become
the longitudinal components of the extra gauge bosons $Z^{\prime
0}$ and $W^{\prime \pm 0}$, giving them masses of order $f \sim 1$
TeV. The gauge and Yukawa couplings break the $SO(5)$ global
symmetry and generate a Coleman-Weinberg type potential
\cite{coleman} for the remaining Goldstone bosons. This potential
gives rise to a vacuum expectation value $v = 174$ GeV for the
complex doublet. However the complex triplet is not protected by
the global symmetry from one loop divergent corrections and hence
obtains a heavy mass of order $f \sim 1$ TeV. Finally, in order to
cancel the quadratic divergence in the top quark one loop
correction to the Higgs mass, it is necessary to introduce a
vector-like (i.e.~non-chiral) weak isosinglet quark T of charge
2/3. As explained in section \ref{protection} vector-like fermions
are not mass protected and so a bare mass term is allowed for the
T quark, which is chosen to be of order $f \sim 1$ TeV.

Little Higgs models have characteristic physics signals at the TeV
scale which may be seen at the LHC. The presence of a vector-like
T quark of charge 2/3 is generically required to cancel the
divergence from the top quark loop. It could be pair produced or
singly together with a jet at hadron colliders, decaying to $Wb$,
$th$ and $tZ$. We also expect new gauge bosons which cancel the
$W$ and $Z$ gauge boson loops. Their quantum numbers, decay modes
and production mechanisms are model dependent. Finally one expects
extra scalars some of which could have a TeV scale mass, like the
triplet in the Littlest Higgs model, and some (typically second
Higgs doublets) could be protected from one loop divergent masses
becoming as light as the SM Higgs. These new states generate
contributions to precision electroweak observables and flavour
changing neutral currents, which provide important constraints on
Little Higgs models.

\subsection{Multiple Point Principle}
\label{mpp1}

In this approach \cite{holger} to the hierarchy problem,
the idea is not to introduce any new physics beyond the SM,
but rather a new fundamental principle: the Multiple Point
Principle (MPP). According to this MPP, nature chooses
coupling constant values such that a number of vacuum states
have the same energy density. This MPP assumption, of course,
explicitly introduces a fine-tuning mechanism, namely the
degeneracy of the vacua. We shall use MPP in section \ref{mpp2}
to predict the top quark and SM Higgs masses, by postulating the
existence of a vacuum state at the fundamental (Planck)
scale $<\phi_{WS}> \sim \Lambda_{Planck} \sim 10^{19}$ GeV
degenerate with the usual SM vacuum having $<\phi_{WS}> = 246$
GeV. This application of the MPP assumes the existence of
the hierarchy $v/\Lambda_{Planck} \sim 10^{-17}$. In order to
actually derive this hierarchy, it is necessary to have
another vacuum at the electroweak scale degenerate with
the above two. This third vacuum is postulated to correspond
to a phase in which an effective 6 $t$ amd 6 $\overline{t}$
S-wave bound state scalar field develops a non-zero vev.
The vacuum degeneracy conditions then predict values for the
top quark Yukawa coupling at both the fundamental scale
$g_t(\mu_{fundamental})$ and the electroweak scale $g_t(v)$.
Now having MPP predicted values for $g_t(\mu)$ at two
different scales, we can calculate the ratio of these
scales from the SM renormalisation group equations (the
SM gauge coupling constants contributing to the running
are considered as given). In this way we can derive the
gauge hierarchy ratio $v/\mu_{fundamental} \sim 10^{-17}$
and show that $\mu_{fundamental} \sim \Lambda_{Planck}
\sim 10^{19}$ GeV. This scenario is discussed in more
detail in Holger Nielsen's lectures \cite{holger}.

\section{Top Quark and Higgs Masses}
\label{top}

\begin{figure}
\leavevmode \vspace{-0.5cm} \centerline{
\epsfig{file=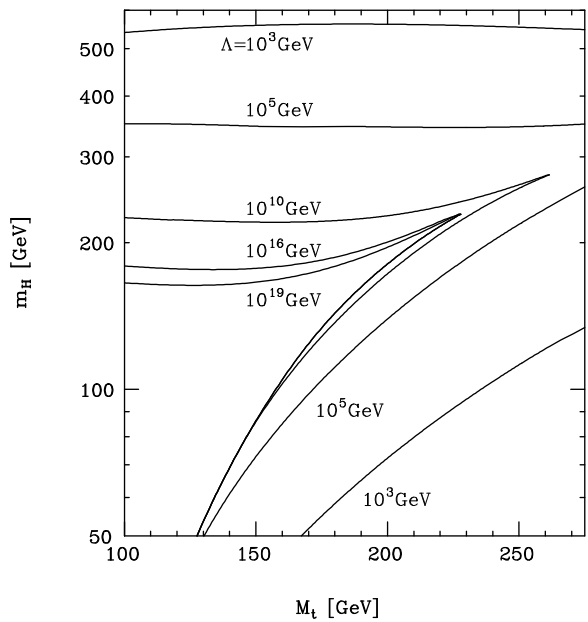,width=10.0cm,angle=90}} \vspace{-1.0cm}
\caption{SM bounds in the ($M_t$,$m_H$) plane for  various values
of $\Lambda$, the scale at which new physics enters.}
\label{fig:Maiani}
\end{figure}

As the top quark mass is the dominant term in the SM fermion mass
matrix, it is likely that its value will be understood dynamically
before those of the other fermions. It has been known for some
time \cite{maiani,sher} that the self-consistency of the pure SM
up to some physical cut-off scale $\Lambda$ imposes constraints on
the top quark and Higgs boson masses. The first constraint is the
so-called triviality bound: the running Higgs coupling constant
$\lambda(\mu)$ should not develop a Landau pole for $\mu <
\Lambda$. The second is the vacuum stability bound: the running
Higgs coupling constant $\lambda(\mu)$ should not become negative
leading to the instability of the usual SM vacuum. These bounds
are illustrated \cite{zwirner} in Fig. \ref{fig:Maiani}, where the
combined triviality and vacuum stability bounds for the SM are
shown for different values of the high energy cut-off $\Lambda$.
The allowed region is the area around the origin bounded by the
co-ordinate axes and the solid curve labelled by the appropriate
value of $\Lambda$. In the following we shall be interested in
large cut-off scales $\Lambda \simeq 10^{15}-10^{19}$ GeV,
corresponding to the grand unified (GUT) or Planck scale. The
upper part of each curve corresponds to the triviality bound. The
lower part of each curve coincides with the vacuum stability bound
and the point in the top right hand corner, where it meets the
triviality bound curve, is the quasi-fixed infra-red fixed point
for that value of $\Lambda$. The vacuum stability curve will be
important for the discussion of the MPP prediction of the top
quark and  SM Higgs boson masses in section \ref{mpp2}. Before
this however, we will discuss their quasi-fixed point values in
the SM and the MSSM.

\subsection{Infra-red Fixed Point}
\label{infrared}

The idea that the top quark mass might be determined dynamically
as an infrared fixed point of the renormalisation group equations
is quite old \cite{pendleton}. In practice one finds an effective
infrared stable quasifixed point at the scale $\mu = m_t$, where
the QCD gauge coupling constant $g_3(\mu)$ is slowly varying
\cite{hill}. The quasifixed point prediction of the top quark mass
is based on two assumptions: (a) the perturbative renormalisation
group equations are valid up to some high (e.~g.~GUT or Planck)
energy scale $\Lambda \simeq 10^{16} - 10^{19}$ GeV, and (b) the
top Yukawa coupling constant is large at the high scale
$g_{t}(\Lambda) \gsim 1$.

Neglecting the lighter quark masses and mixings, which is a good
approxmation, the SM one loop renormalisation group equation for
the top quark Yukawa coupling $g_t(\mu) = \sqrt{2}
m_t(\mu)/\langle\phi_{WS}\rangle$ is:
\begin{equation}
16\pi^2\frac{dg_t}{d\ln\mu} = g_t\left(\frac{9}{2}g_t^2 - 8g_3^2 -
\frac{9}{4}g_2^2 - \frac{17}{12}g_1^2\right)
\end{equation}
The gauge coupling constants $g_i(\mu)$ satisfy the equations:
\begin{equation}
 16\pi^2\frac{dg_i}{d\ln\mu} = b_ig_i^3  \quad \rm{where} \quad b_1
 = \frac{41}{6}, \quad b_2 = -\frac{19}{6}, \quad b_3 =-7 \quad
 \rm{in \; the \; SM.}
 \label{rgegauge}
\end{equation}
The nonlinearity of the renormalisation group equations then
strongly focuses $g_{t}(\mu)$ at the electroweak scale to its
quasifixed point value. We note that while there is a rapid
convergence to the top Yukawa coupling fixed point value from
above, the approach from below is much more gradual. The
renormalisation group equation for the Higgs self-coupling
$\lambda(\mu) = m_H^2(\mu)/\langle\phi_{WS}\rangle^2$
\begin{equation}
 16\pi^2\frac{d\lambda}{d\ln\mu} =12\lambda^2 + 3\left(4g_t^2 -
 3g_2^2 - g_1^2\right)\lambda + \frac{9}{4}g_2^4 +
 \frac{3}{2}g_2^2g_1^2 + \frac{3}{4}g_1^4 - 12g_t^4
 \label{rgelam}
\end{equation}
similarly focuses $\lambda(\mu)$ towards a quasifixed point value,
leading to the SM fixed point predictions \cite{hill} for the
running top quark and Higgs masses:
\begin{equation}
m_{t} \simeq 225\ \mbox{GeV} \quad m_{H} \simeq 250\ \mbox{GeV}
\end{equation}
Unfortunately these predictions are inconsistent with the FermiLab
results, which require a running top mass \mbox{$m_{t} \simeq
166 \pm 5$ GeV}.

\begin{figure}
\leavevmode \centerline{ \epsfxsize=6.75cm \epsfbox{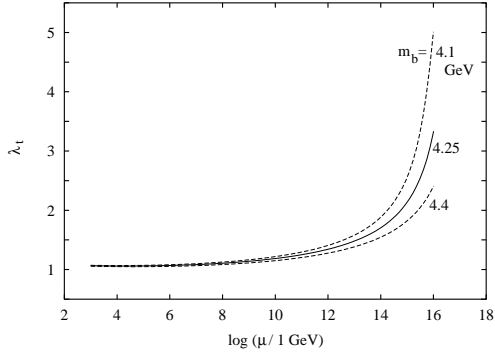}}
\caption{The rapid convergence from above of the top Yukawa
coupling constant $\lambda_t \equiv g_t$ to the MSSM infra-red
quasi-fixed point value as $\mu \rightarrow m_t$.}
\label{fig:bargfp}
\end{figure}

The fixed point top Yukawa coupling is reduced by about 15\% in
the MSSM to a value of $g_t(m_t) \simeq 1.1$ as shown in Fig.
\ref{fig:bargfp} \cite{barger}, due to the contribution of the
superpartners to the renormalisation group beta functions.
Also the top quark couples to just one of the MSSM Higgs doublets
$H_U$, leading to the fixed point prediction
\begin{equation}
 m_t(m_t) \simeq (190\ \mbox{GeV})\sin\beta
 \label{mssmfp}
\end{equation}
for $\tan\beta\lsim 15$. Including the stop/gluino one-loop SUSY
correction the experimental running top quark mass becomes
$m_t = 160 \pm 5$ GeV. Thus agreement with the fixed point
prediction  would require $\tan\beta =1.5 \pm 0.3$, which is
ruled out in the MSSM at the $3\sigma$ level by LEP data \cite{pdg}.
However we note that the infrared fixed point scenerio for the
top quark mass is consistent in the next to minimal supersymmetric
standard model (NMSSM) \cite{roman}.

For large $\tan\beta$ it is possible to have a bottom quark Yukawa
coupling satisfying \mbox{$g_{b}(\Lambda) \gsim 1$} which then
approaches an infrared quasifixed point and is no longer
negligible in the renormalisation group equation for $g_{t}(\mu)$.
Indeed with $\tan\beta \simeq 60$, we can trade the mystery of the
top to bottom quark mass ratio for that of a hierarchy of vacuum
expectation values, \mbox{$v_{2}/v_{1} \simeq 60$}, and have all
the third generation Yukawa coupling constants large:
\begin{equation}
g_{t}(\Lambda) \gsim 1 \quad g_{b}(\Lambda) \gsim 1 \quad
g_{\tau}(\Lambda) \gsim 1 \label{tbtaufp}
\end{equation}
Then $m_{t}$, $m_{b}$ and $m_{\tau}$ all approach infrared
quasifixed point values\footnote{However we note, in the large
$\tan\beta$ scenario, SUSY radiative corrections to $m_{b}$ are
generically large: the bottom quark mass gets a contribution
proportional to $v_{2}$ from some one-loop diagrams,
%with internal superpartners, such as top squark-charged Higgsino exchange
whereas its tree level mass is proportional to $v_{1} =
v_{2}/\tan\beta$.} compatible with experiment \cite{fkm,kazakov1}.
This large $\tan\beta$ scenario is consistent with the idea of
Yukawa unification \cite{anan,kazakov2}:
\begin{equation}
g_{t}(\Lambda) = g_{b}(\Lambda) = g_{\tau}(\Lambda) = g_{G}
\label{yukun}
\end{equation}
as occurs in the SO(10) SUSY-GUT model with the two MSSM Higgs
doublets in a single {\bf 10} irreducible representation and
$g_{G} \gsim 1$ ensures fixed point behaviour. However it should
be noted that the equality in eq.~(\ref{yukun}) is not necessary
to obtain the fixed point values. Also the lightest Higgs particle
mass is predicted to be $m_{h^0} \simeq 120$ GeV.

\subsection{Multiple Point Principle}
\label{mpp2}

\begin{figure}
\leavevmode \vspace{-0.5cm}
\centerline{\epsfig{file=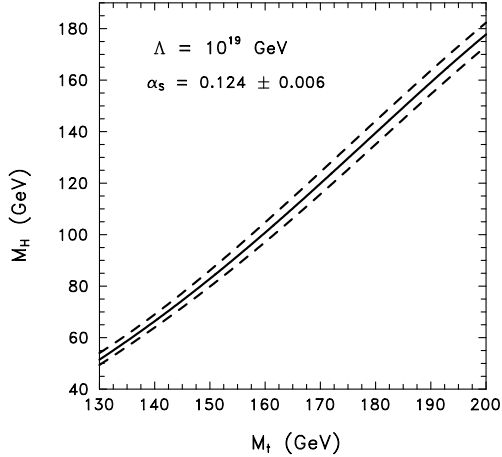,width=7cm,%
bbllx=95pt,bblly=130pt,bburx=510pt,bbury=555pt,%
angle=270,clip=} } \caption{SM vacuum stability curve for $\Lambda
= 10^{19}$ GeV and $\alpha_s = 0.124$ (solid line), $\alpha_s =
0.118$ (upper dashed line), $\alpha_s = 0.130$ (lower dashed
line).} \label{fig:vacstab}
\end{figure}

The application of the MPP to the pure Standard Model \cite{fn2},
with a cut-off close to $\Lambda_{Planck}$, implies that the SM
parameters should be adjusted, such that there exists another
vacuum state degenerate in energy density with the vacuum in which
we live. This means that the effective SM Higgs potential
$V_{eff}(\phi_{WS})$ should have a second minimum degenerate with
the well-known first minimum at the electroweak scale $<\phi_{WS}>
= \phi_{vac\; 1} = 246$ GeV. Thus we predict that our vacuum is
barely stable and we just lie on the vacuum stability curve in the
top quark, Higgs particle (pole) mass ($M_t$, $M_H$) plane, shown
\cite{casas} in Fig. \ref{fig:vacstab} for a cut-off $\Lambda =
10^{19}$ GeV . Furthermore we expect the second minimum to be
within an order of magnitude or so of the fundamental scale, i.e.
$<\phi_{WS}> = \phi_{vac\; 2} \simeq \Lambda_{Planck}$. In this
way, we essentially select a particular point on the SM vacuum
stability curve and hence the MPP condition predicts precise
values for $M_t$ and $M_H$.

For the purposes of our discussion it is sufficient to consider
the renormalisation group improved tree level effective potential
$V_{eff}(\phi_{WS})$. We are interested in values of the Higgs
field of the order $\phi_{vac\; 2} \simeq \Lambda_{Planck}$, which
is very large compared to the electroweak scale, and for which the
quartic term strongly dominates the $|\phi_{WS}|^2$ term; so to a
very good approximation we have:
\begin{equation}
V_{eff}(\phi_{WS}) \simeq \frac{1}{8}\lambda (\mu = |\phi_{WS} | )
|\phi_{WS} |^4
\end{equation}
The running Higgs self-coupling constant $\lambda (\mu)$ and the
top quark running Yukawa coupling constant $g_t(\mu)$ are readily
computed by means of the renormalisation group equations, which
are in practice solved numerically, using the second order
expressions for the beta functions.

\begin{figure}
\leavevmode
\centerline{
\epsfig{file=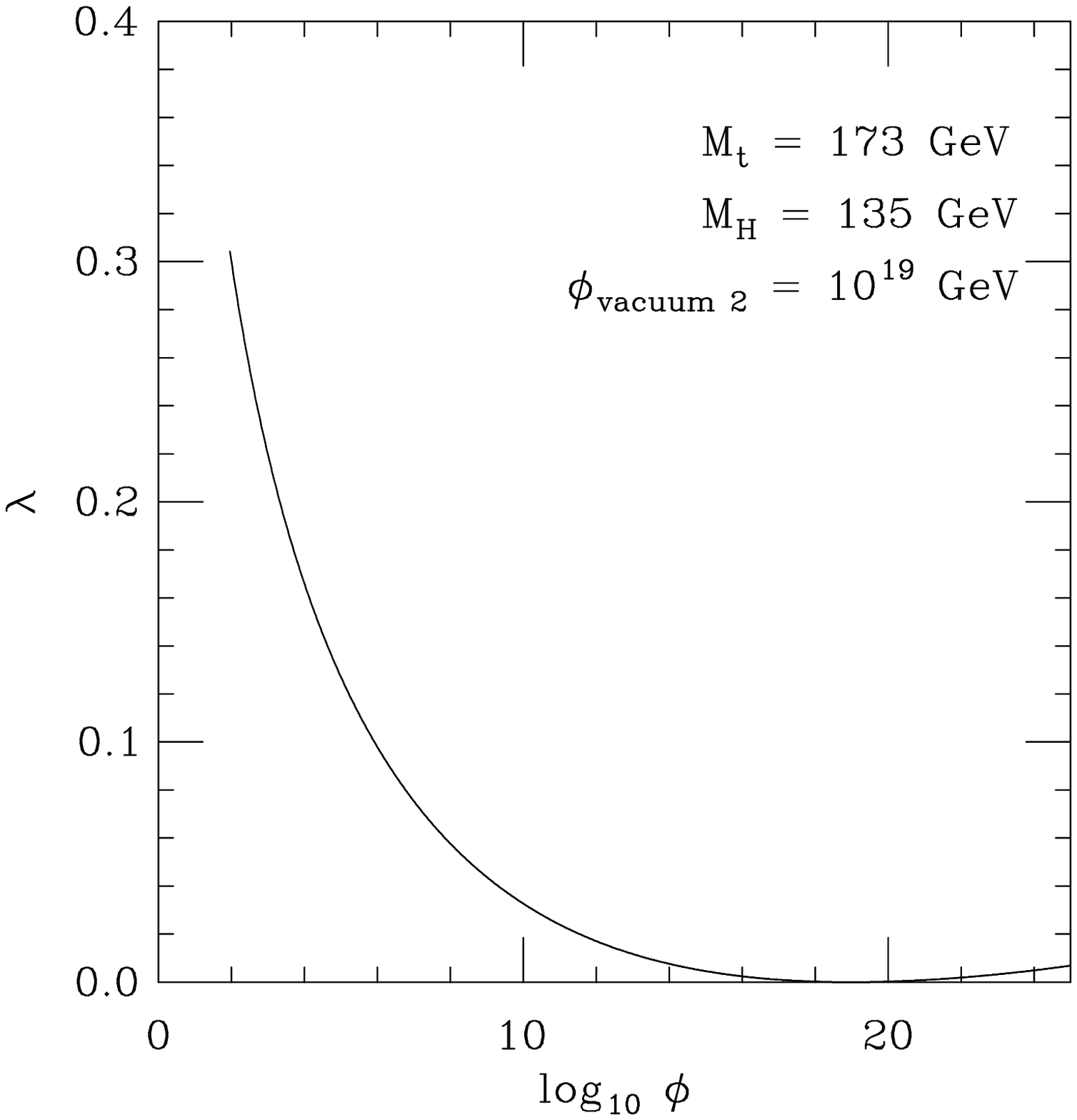,width=6.8cm}
\hspace{-0.6cm}
\epsfig{file=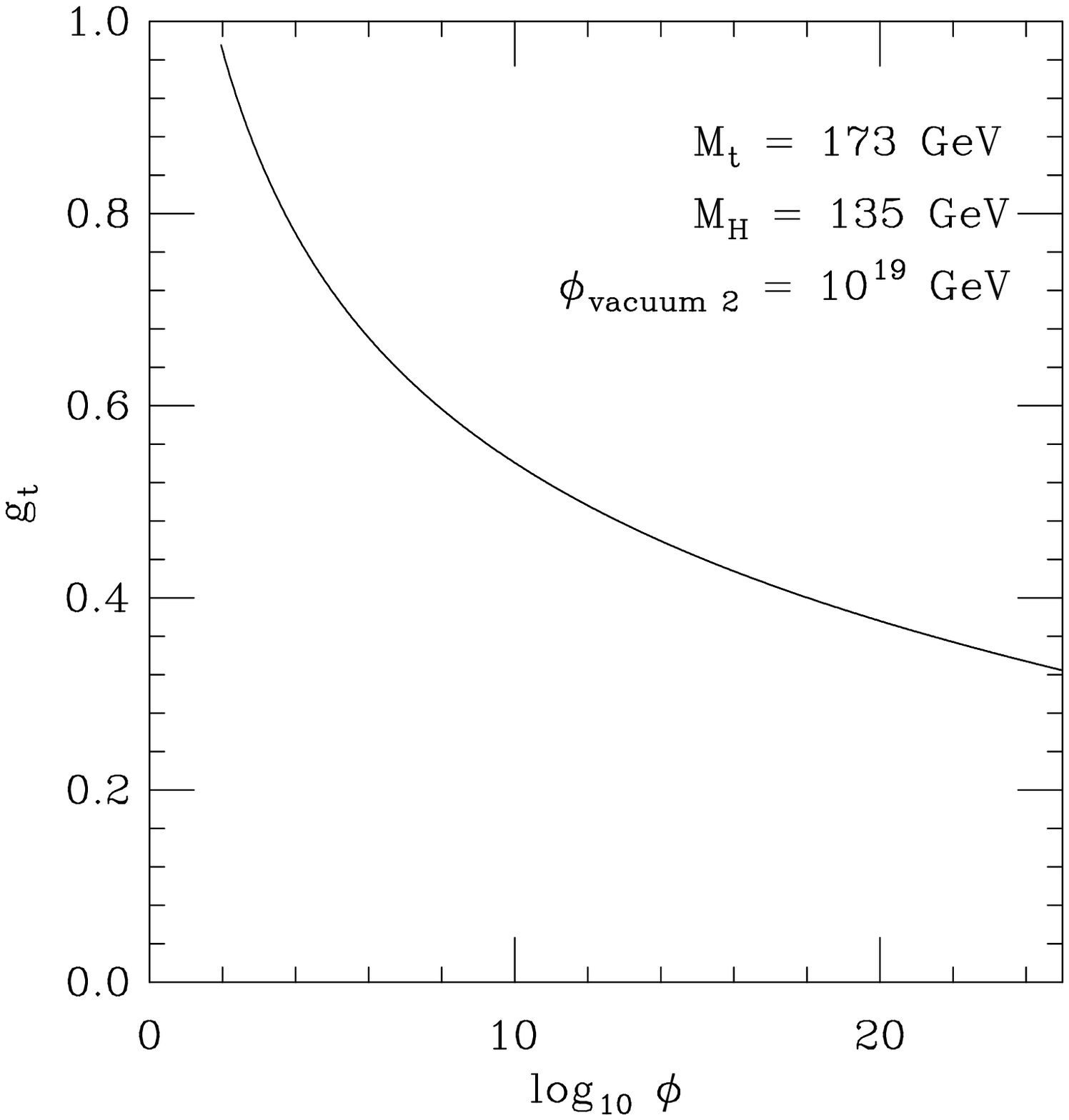,width=6.8cm}
}
\vspace{-0.6cm}
\caption{Plots of $\lambda$
and $g_t$ as functions
of the scale of the
Higgs field $\phi$ for degenerate vacua with the second Higgs
VEV at the Planck scale $\phi_{vac\;2}=10^{19}$ GeV.
We formally apply the second order SM renormalisation
group equations up to a scale of $10^{25}$ GeV.}
\label{fig:lam19}
\end{figure}

The vacuum degeneracy condition is imposed by requiring:
\begin{equation}
V_{eff}(\phi_{vac\; 1}) = V_{eff}(\phi_{vac\; 2}) \label{eqdeg}
\end{equation}
Now the energy density in vacuum 1 is exceedingly small compared
to $\phi_{vac\; 2}^4 \simeq \Lambda_{Planck}^4$. So we basically
get the degeneracy condition, eq.~(\ref{eqdeg}), to mean that the
coefficient $\lambda(\phi_{vac\; 2})$ of $|\phi_{vac\; 2}|^4$ must
be zero with high accuracy. At the same $\phi$-value the
derivative of the effective potential $V_{eff}(\phi_{WS})$ should
be zero, because it has a minimum there. Thus at the second
minimum of the effective potential the beta function
$\beta_{\lambda}$ also vanishes:
\begin{equation}
\beta_{\lambda}(\mu = \phi_{vac\; 2}) = \lambda(\phi_{vac\; 2}) =
0
\end{equation}
which gives to leading order the relationship:
\begin{equation}
\frac{9}{4}g_2^4 + \frac{3}{2}g_2^2g_1^2 + \frac{3}{4}g_1^4 -
12g_t^4 = 0
\end{equation}
between the top quark Yukawa coupling and the electroweak gauge
coupling constants $g_1(\mu)$ and $g_2(\mu)$ at the scale $\mu =
\phi_{vac\; 2} \simeq \Lambda_{Planck}$. We use the
renormalisation group equations to relate the couplings at the
Planck scale to their values at the electroweak scale. Figure
\ref{fig:lam19} shows the running coupling constants
$\lambda(\phi)$ and $g_t(\phi)$ as functions of $\log(\phi)$.
Their values at the electroweak scale give our predicted
combination of pole masses \cite{fn2}:
\begin{equation}
M_{t} = 173 \pm 5\ \mbox{GeV} \quad M_{H} = 135 \pm 9\ \mbox{GeV}
\end{equation}

\section{Mass Matrix Ans\"{a}tze and Texture}
\label{ansatze}

The motivation for considering mass matrix ans\"{a}tze is to
obtain testable relationships between fermion masses and mixing
angles, which might reduce the number of parameters in the Yukawa
sector and provide a hint to the physics beyond the SM. We shall
focus on the charged fermion sector here, particularly the quarks,
and postpone discussion of neutrino masses and mixings to section
\ref{neutrino}. The hierarchical structure of the mass spectrum
should be reflected in the fermion mass matrices. So it is natural
to speculate that the smaller matrix elements may contribute so
weakly to the physical masses and mixing angles that they can
effectively be neglected and replaced by zero---the so-called
texture zeros. The most famous ansatz incorporating such a texture
zero is the Fritzsch hermitian ansatz \cite{fritzsch} for the two
generation quark mass matrices:
%.
\begin{equation}
M_U =\pmatrix{0      & B\cr
          B^{\ast}   & A\cr}
\qquad
M_D =\pmatrix{0          & B^\prime\cr
          B^{\prime\ast}  & A^\prime\cr}
\end{equation}
The assumed hierarchical structure gives the following conditions:
\begin{equation}
|A| \gg |B| , \qquad |A^\prime| \gg |B^\prime|
\end{equation}
among the parameters.
It follows that the two generation
Cabibbo mixing is given by the well-known
Fritzsch formula
\begin{equation}
\left| V_{us} \right| \simeq
\left|\sqrt{\frac{m_d}{m_s}} -
e^{i\phi} \sqrt{\frac{m_u}{m_c}} \right|
\label{fritzsch1}
\end{equation}
where $\phi = \arg{B^\prime} - \arg{B}$. This relationship
fits the experimental value well, provided
that the phase $\phi$ is close to $\frac{\pi}{2}$.
The generalisation of the Fritzsch ansatz to three generations:
%.
\begin{equation}
M_U =\pmatrix{0         & C         & 0\cr
              C^\ast        & 0         & B\cr
              0         & B^\ast        & A\cr}
%\end{equation}
%\begin{equation}
\qquad
M_D =\pmatrix{0         & C^\prime  & 0\cr
              C^{\prime\ast}    & 0         & B^\prime\cr
              0         & B^{\prime\ast}  & A^\prime\cr}
\end{equation}
with the assumed hierarchy of parameters:
\begin{equation}
|A| \gg |B| \gg |C|, \qquad |A^\prime| \gg |B^\prime| \gg |C^\prime|
\end{equation}
however leads to an additional relationship
\begin{equation}
|V_{cb}| \simeq
\left| \sqrt{\frac{m_{s}}{m_{b}}} -
e^{-i\phi_{2}}\sqrt{\frac{m_{c}}{m_{t}}} \right|
\label{fritzsch2}
\end{equation}
which is excluded by the data for any value of
the phase $\phi_2$.
Consistency with experiment can, for example, be restored by
introducing a non-zero $(M_U)_{22}$ mass matrix element \cite{xing}.

Several ans\"{a}tze have been proposed---for example, see
\cite{rrr} for a systematic analysis of symmetric quark mass
matrices with texture zeros at the SUSY-GUT scale. Here I will
concentrate on the lightest family mass generation model
\cite{lfm}, which successfully expresses all the quark mixing
angles in terms of simple, compact formulae involving quark mass
ratios. According to this model the flavour mixing for quarks is
basically determined by the mechanism responsible for generating
the physical masses of the up and down quarks, $m_{u}$ and $m_{d}$
respectively. So, in the chiral symmetry limit, when $m_{u}$ and
$m_{d}$ vanish, all the quark mixing angles vanish. Therefore we
are led to consider an ansatz in which the diagonal mass matrix
elements for the second and third generations are practically the
same in the gauge (unrotated) and physical bases.

The mass matrix for the down quarks ($D$ = $d$, $s$, $b$) is taken
to be hermitian with three texture zeros of the following form:
\begin{equation}
M_{D}=\pmatrix{ 0 & a_D & 0 \cr a_D^{\ast} & A_D & b_D \cr 0 &
b_D^{\ast} & B_D \cr}  \label{LFM1}
\end{equation}
where
\begin{equation}
B_{D}=m_{b}+\delta_{D} \qquad A_{D}= m_{s} + \delta_{D}^{\prime }
\qquad |\delta_D |\ll m_{s} \qquad |\delta _{D}^{\prime }|\ll m_{d}
\label{BA}
\end{equation}
It is, of course, necessary to assume some hierarchy between the
elements, which we take to be: $B_{D}\gg A_{D}\sim \left|
b_{D}\right| \gg \left| a_{D}\right| $. The zero in the $\left(
M_{D}\right) _{11}$ element corresponds to the commonly accepted
conjecture that the lightest family masses appear as a direct
result of flavour mixings. The zero in $\left( M_{D}\right) _{13}$
means that only minimal ``nearest neighbour'' interactions occur,
giving a tridiagonal matrix structure. Since the trace and
determinant of the hermitian matrix $M_{D}$ gives the sum and
product of its eigenvalues, it follows that
\begin{equation}
\delta _{D}\simeq - m_{d}  \label{del}
\end{equation}
while $\delta _{D}^{\prime }$ is vanishingly small and can be neglected
in further considerations.

It may easily be shown that equations (\ref{LFM1} - \ref{del}) are
entirely equivalent to the condition that the diagonal
elements ($A_{D}$, $B_{D}$) of $M_{D}$ are proportional
to the modulus square of the off-diagonal elements ($a_{D}$, $b_{D}$):
\begin{equation}
\frac{A_{D}}{B_{D}}=\left| \frac{a_{D}}{b_{D}}\right| ^{2}
\label{ABab}
\end{equation}
Using the conservation of the trace, determinant and sum of
principal minors of the hermitian matrix $M_{D}$ under unitary
transformations, we are led to a complete determination of the
moduli of all its elements, which can be expressed to high
accuracy as follows:
\begin{equation}
\left| M_{D} \right| = \pmatrix{ 0 & \sqrt{m_d m_s} & 0 \cr
\sqrt{m_d m_s} & m_s & \sqrt{m_d m_b} \cr 0 &
\sqrt{m_d m_b} & m_b - m_d \cr}  \label{LFM1A}
\end{equation}

The mass matrix for the up quarks is taken to be of the following
hermitian form:
\begin{equation}
M_{U}=\pmatrix{ 0 & 0 & c_U \cr 0 & A_U & 0 \cr c_U^{\ast} & 0 & B_ U \cr}
\label{LFM2}
\end{equation}
The moduli of all the elements of $M_{U}$ can also be readily
determined in terms of the physical masses as follows:
\begin{equation}
\left| M_{U} \right| = \pmatrix{ 0 & 0 & \sqrt{m_u m_t} \cr
0 & m_c & 0 \cr \sqrt{m_u m_t} & 0 & m_t - m_u \cr}
\label{LFM2A}
\end{equation}

The CKM quark mixing matrix elements can now be readily calculated
by diagonalising the mass matrices $M_D$ and $M_U$. They are
given  in terms of quark mass ratios as follows:
\begin{eqnarray}
\left| V_{us}\right| = \sqrt{\frac{m_{d}}{m_{s}}} = 0.222 \pm 0.004
\qquad \left| V_{us}\right|_{exp} = 0.221 \pm 0.003 \\
\left|V_{cb}\right| = \sqrt{\frac{m_{d}}{m_{b}}} = 0.038 \pm 0.004
\qquad \left|V_{cb}\right|_{exp} = 0.039 \pm 0.003 \\
\left|V_{ub}\right| = \sqrt{\frac{m_{u}}{m_{t}}} = 0.0036 \pm 0.0006
\qquad \left|V_{ub}\right|_{exp} = 0.0036 \pm 0.0006   \label{angles}
\end{eqnarray}
As can be seen, they are in impressive agreement with the experimental
values.

The proportionality condition (\ref{ABab}) is not so easy to
generate from an underlying symmetry beyond the Standard Model,
but it is possible to realise it in a local chiral $SU(3)$ family
symmetry model \cite{SU3}.

It is common to make ans\"{a}tze in the context of SUSY-GUT models,
which of course give relationships between the fermion mass
parameters at the grand unified scale $\Lambda_{GUT}$. In grand
unified theories, the SM gauge group
$SMG \equiv SU(3) \times SU(2) \times U(1)$ is embedded in a
simple Lie group $G$ and the SM interactions become unified
for scales $\mu > \Lambda_{GUT}$. Also each generation of
quark and lepton SM irreducible representations are combined
into larger irreducible representations of $G$, introducing
symmetries between quarks and leptons. In the minimal $SU(5)$
model the two MSSM Higgs doublets are promoted
into ${\bf 5}$ and $\mathbf{ \overline{5}}$ representations,
giving the well-known GUT relation
\begin{equation}
m_b(\Lambda_{GUT}) = m_{\tau}(\Lambda_{GUT}),
\label{mbmtau}
\end{equation}
which is consistent with experiment. However the model also predicts
the equality of the full down quark and charged lepton mass
matrices ($M_D = M_E$) and, in particular:
\begin{equation}
m_{d}/m_{s} = m_{e}/m_{\mu}
\label{eq:massratio}
\end{equation}
which fails phenomenologically by an order of magnitude. So it is
necessary to introduce a more complicated group theoretical structure.

Georgi and Jarlskog \cite{georgijarlskog} introduced a Higgs field
in the the 45 dimensional representation of $SU(5)$ and postulated
that the Weinberg Salam Higgs field $H_D$ resides in a linear
combination of the $\mathbf{\overline{5}}$ and
$\mathbf{\overline{45}}$ representations. Furthermore they assumed
the following ansatz, in which the $\mathbf{\overline{45}}$ only
contributes to the $(M_D)_{22}$ and $(M_E)_{22}$ matrix elements 
and dominates them:
\begin{equation}
M_D =\pmatrix{0          & F        & 0\cr
          F              & E        & 0\cr
          0              & 0        & D\cr}
\quad
M_E =\pmatrix{0         & F         & 0\cr
          F         & -3E       & 0\cr
          0         & 0         & D\cr}
\label{eq:gjansatz}
\end{equation}
Then, with the assumed hierarchical parameters
$|D| \gg |E| \gg |F|$, we are led to the mass relations
$m_{b}(\Lambda_{GUT}) = m_{\tau}(\Lambda_{GUT})$,
$m_{s}(\Lambda_{GUT}) = m_{\mu}(\Lambda_{GUT})$/3 and
$m_{d}(\Lambda_{GUT}) = 3m_{e}(\Lambda_{GUT})$,
which are consistent with experiment.
This ansatz is easily extended \cite{harvey} to $SO(10)$, in
which right-handed neutrinos are introduced for each generation.
There are then 16 fermion states for each generation,
which fit into the single {\bf 16} representation of $SO(10)$
and are such that the sum of the $(B-L)$ charges for the 16 fermions
vanishes. So the $(B-L)$ charge can be gauged and
$SU(5) \times U(1)_{B-L}$ is a subgroup of $SO(10)$.
However, in order to also generate realistic up quark masses and
CKM matrix, it is necessary to further complicate the group
theoretical ansatz, see for example \cite{anderson,blazek,matsuda}.

We now turn to the question of the dynamics underlying the
hierarchical texure of the above ans\"{a}tze.

\section{Origin of Texture}
\label{origin}

As pointed out in section \ref{protection}, a natural explanation
of the charged fermion mass hierarchy would be mass protection,
due to the existence of some approximately conserved chiral
charges beyond the SM. These chiral charges, which we should like
to identify with gauge quantum numbers in the fundamental theory
beyond the SM, provide selection rules forbidding the transitions
between the various left-handed and right-handed quark-lepton
states, except for the top quark. In order to generate mass terms
for the other fermion states, we have to introduce new Higgs
fields, which break the symmetry group $G$ of the fundamental
theory down to the SM group. We also need suitable intermediate
fermion states to mediate the forbidden transitions, which we take
to be vector-like Dirac fermions with a mass of order the
fundamental scale $M_F$ of the theory. In this way effective SM
Yukawa coupling constants are generated \cite{fn1}, which are
suppressed by the appropriate product of Higgs field vacuum
expectation values measured in units of $M_F$. We assume that all
the couplings in the fundamental theory are unsuppressed,
i.e.~they are all naturally of order unity.

Consider, for example, the model obtained by extending the
Standard Model gauge group $SMG = SU(3) \times SU(2) \times U(1)$
with a gauged chiral abelian flavour group $U(1)_f$. This $SMG \times
U(1)_f$ gauge group is assumed to be broken to $SMG$ by the vev of
a scalar field $\phi_S$ where $\langle\phi_S\rangle <  M_F$ and
$\phi_S$ carries $U(1)_f$ charge $Q_f(\phi_S)$ = 1. Suppose
further that the $U(1)_f$ charges of the Weinberg Salam Higgs
field and the left- and right-handed bottom quark fields are:
\begin{equation}
Q_f(\phi_{WS})=0 \qquad Q_f(b_L)=0 \qquad Q_f(b_R)=2
\end{equation}
Then it is natural to expect the generation of a mass for the $b$
quark of order:
\begin{equation}
m_b \simeq \left( \frac{<\phi_S> }{M_F} \right)^2<\phi_{WS}>
\end{equation}
via the tree level diagram shown in Fig. \ref{fig:pic}, involving
the exchange of two $\langle\phi_S\rangle$ tadpoles, in addition
to the usual $\langle\phi_{WS}\rangle$ tadpole, with two
appropriately charged vector-like fermion intermediate states of
mass $M_F$. We identify $\epsilon_f=\langle\phi_S\rangle/M_F$ as
the $U(1)_f$ flavour symmetry breaking parameter. In general we
expect mass matrix elements of the form:
\begin{equation}
M(i,j) = \gamma_{ij} \epsilon_{f}^{n_{ij}}\langle\phi_{WS}\rangle
\label{eq:mij}
\end{equation}
between the $i$th left-handed and $j$th right-handed fermion
components, where
\begin{equation}
\gamma_{ij} = {\cal O} (1), \quad n_{ij}= \mid Q_f(\psi_{L_{i}}) -
Q_f(\psi_{R_{j}})\mid \label{eq:gammaij}
\end{equation}
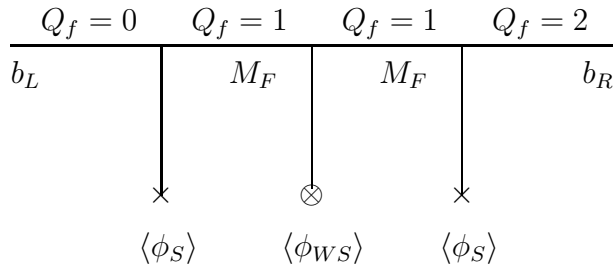
\begin{figure}[hbt]
\begin{center}
\setlength{\unitlength}{1mm}
\begin{picture}(80,40)(0,15)
\put(0,45){\line(1,0){80}} \put(20,45){\line(0,-1){20}}
\put(40,45){\line(0,-1){19}} \put(60,45){\line(0,-1){20}}
\put(18.3,24){$\times$} \put(38.3,24){$\otimes$}
\put(58.3,24){$\times$} \put(17,17){$\langle\phi_{S}\rangle$}
\put(36,17){$\langle\phi_{WS}\rangle$}
\put(57,17){$\langle\phi_{S}\rangle$} \put(4,47){$Q_f = 0$}
\put(24,47){$Q_f = 1$} \put(44,47){$Q_f = 1$} \put(64,47){$Q_f =
2$} \put(0,40){$\large{b}_L$} \put(76,40){$\large{b}_R$}
\put(29,40){$M_F$} \put(49,40){$M_F$}
\end{picture}
\end{center}
\par
\caption{Feynman diagram which generates the b quark mass via
superheavy intermediate states.}
\label{fig:pic}
\end{figure}
So the {\em effective} SM Yukawa couplings of the quarks and
leptons to the Weinberg-Salam Higgs field $y_{ij} =
\gamma_{ij}\epsilon_{f}^{n_{ij}}$ can consequently be small even
though all {\em fundamental\/} Yukawa couplings of the ``true''
underlying theory are of $\cal O$(1). We are implicitly assuming
here that there exists a superheavy spectrum of states which can
mediate all of the symmetry breaking transitions; in particular we
do not postulate the {\em absence} of appropriate superheavy
states in order to obtain exact texture zeros in the mass
matrices. The mass matrix elements, eq. (\ref{eq:mij}), are only
predicted in order of magnitude, unless one can specify the
spectrum of superheavy fermions and their Yukawa couplings.

It does not seem possible to explain the SM fermion mass spectrum
with an anomaly free set of flavour charges in an $SMG \times
U(1)_f$ model with a single Higgs field $\phi_S$ breaking the
$U(1)_f$ gauge symmetry. One can introduce a second gauged
abelian\footnote{Non-abelian $SU(2)$ and $SU(3)$ flavour symmetry
groups have also been considered, see for example
\cite{SU3,raby,SU3ross} for some recent examples.} flavour group
$U(1)_{f^{\prime}}$ and obtain a realistic charged fermion
spectrum in an $SMG \times U(1)_f \times U(1)_{f^{\prime}}$
model\footnote{This $SMG \times U(1)^2$ model has exactly the same
charged fermion spectrum as the so-called anti-grand unification
model \cite{smg3m}.} \cite{smgu2}. In supersymmetric models one
can allow an abelian flavour gauge symmetric extension of the MSSM
to be anomalous \cite{ibanezross,irges}, this anomaly being
cancelled by the Green-Schwartz mechanism \cite{green-schwarz}.
Models with this property clearly point towards a string theory
origin.

The texture of some of the fermion mass matrix ans\"{a}tze in
SUSY-GUT models, such as \cite{anderson,blazek}, can be generated
from global chiral abelian flavour models possibly combined with
some discrete symmetries. However the required spectrum of Higgs
fields and superheavy matter fields tends to be rather complicated.
Texture zeros appear as a consequence of the absence of the
superheavy states needed to mediate the transition between the
corresponding SM Weyl states in diagrams similar to
Fig. \ref{fig:pic}. For example an explicit $SO(10)$ SUSY-GUT model
has been constructed by Albright and Barr \cite{albright}, in which
the $SO(10)$ and mass protecting flavour quantum numbers of all the
Higgs and matter fields are specified. It is based on a global
$U(1) \times Z_2 \times Z_2$ flavour symmetry that stabilizes a
solution to the doublet-triplet splitting problem, mentioned in
section \ref{susy}, and determines the structure of the Higgs and
Yukawa superpotentials from which the fermion mass matrices are
constructed. The required $SO(10)$ representations of Higgs fields are:
one {\bf 45}, two $\mathbf{16} \oplus \mathbf{\overline{16}}$ pairs,
six {\bf 10} and five {\bf 1}. In addition to the three {\bf 16}
representations for the the quarks and leptons, the superheavy
matter fields comprise the following $SO(10)$ representations:
two $\mathbf{16} \oplus \mathbf{\overline{16}}$ pairs, two {\bf 10}
and six {\bf 1}. Finally to avoid too rapid proton decay, there
is also a discrete $Z_2$ matter symmetry. The resulting mass matrices
for the down quarks and leptons are
\begin{equation}
M_D =\pmatrix{\eta       &\delta  &\delta^{\prime}e^{i\phi}\cr
          \delta         & 0           &\sigma + \epsilon/3\cr
          \delta^{\prime}e^{i\phi}  &-\epsilon/3     & 1\cr}m_b^0
\qquad
M_E =\pmatrix{\eta     & \delta    &\delta^{\prime}e^{i\phi}\cr
      \delta               & 0            & -\epsilon\cr
      \delta^{\prime}e^{i\phi} &\sigma + \epsilon &1\cr}m_b^0
\label{eq:ab1}
\end{equation}
and the up quark and Dirac neutrino\footnote{$M_N$ is inserted in
the well-known see-saw formula for the light neutrino mass matrix
given in section \ref{neutrino}.} mass matrices are:
\begin{equation}
M_U =\pmatrix{\eta       & 0                 & 0\cr
               0         & 0            & \epsilon/3 \cr
               0         & -\epsilon/3        & 1\cr}m_t^0
\qquad
M_N =\pmatrix{\eta       & 0                 & 0\cr
               0         & 0            & \epsilon \cr
               0         & -\epsilon        & 1\cr}m_t^0
\label{eq:ab2}
\end{equation}
where there is a natural hierarchy of parameters $\sigma \sim 1
\gg \epsilon \gg \delta, \delta^\prime \gg\eta$. Since the
spectrum of heavy states has been fully specified, there are no
missing $\mathcal{O}(1)$ factors in eqs. \ref{eq:ab1} and
\ref{eq:ab2}. The nine SM charged fermion masses, the three CKM
mixing angles and CP violating phase are well-fitted with the 8
parameters in the above matrices, after renormalisation group
evolution from the GUT scale.

The flavour symmetry group should of course be a natural feature
of the fundamental theory beyond the SM. In the case of SUSY-GUT
models, it is hoped that the group arises from string theory. An
attractive possibility is that the flavour symmetry group is simply
a sub-group of the gauge symmetry group of the fundamental theory.
This is the case in the so-called anti-grand unification theory
(AGUT) or family replicated gauge group model \cite{smg3m,fnt}.
The AGUT model is based on a non-supersymmetric non-simple
extension of the SM with three copies of the SM gauge group---one
for each family or generation. With the inclusion of three
right-handed neutrinos, the AGUT gauge group becomes $G = (SMG
\times U(1)_{B-L})^3$, where the three copies of the SM gauge
group are supplemented by an abelian $(B-L)$ (= baryon number
minus lepton number) gauge group for each family\footnote{The
family replicated gauge groups $(SO(10))^3$ and $(E_6)^3$ have
recently been considered by Ling and Ramond \cite{ling}.}. The
AGUT gauge group $G$ is the largest anomaly free group,
transforming the known 45 Weyl fermions plus the three
right-handed neutrinos into each other unitarily, which does {\em
not} unify the irreducible representations under the SM gauge
group. It is supposed to be effective at energies near to the
Planck scale, where the $i$'th proto-family couples to just the
$i$'th $SMG\times U(1)_{B-L}$ factor. This gauge group is broken
down by four Higgs fields $W$, $T$, $\rho$ and $\omega$, having
vevs about one order of magnitude lower than the Planck scale, to
its diagonal subgroup:
\begin{equation}
(SMG \times U(1)_{B-L})^3 \rightarrow SMG\times U(1)_{B-L}
\end{equation}
The diagonal $U(1)_{B-L}$ is broken down at the see-saw scale by
another Higgs field $\phi_{SS}$ and the diagonal $SMG$ is broken
down to $SU(3) \times U(1)_{em}$ by the Weinberg-Salam Higgs
field. We note that the $(SMG \times U(1)_{B-L})^3$ gauge quantum
numbers of the quarks and leptons are uniquely determined by the
structure of the model and they include 6 chiral abelian
charges---the weak hypercharge and (B-L) quantum number for each
of the three generations. With an appropriate choice of the
abelian charges of the Higgs fields, it is possible to generate a
good order of magnitude fit to the SM fermion masses, as discussed
in Holger's lectures \cite{holger}. In this fit, we do not attempt
to guess the spectrum of superheavy fermions at the Planck scale,
but simply assume a sufficiently rich spectrum to mediate all of
the symmetry breaking transitions in the various mass matrix
elements.

\section{Neutrino Mass and Mixing}
\label{neutrino}

The recent impressive experimental developments in neutrino
physics have been described here by John Simpson \cite{simpson}
and the formalism of neutrino oscillations has been described by
Boris Kayser \cite{kayser}. The phenomenology of neutrino
oscillations \cite{garcia} is summarised in the neutrino mass
squared differences of eq. (\ref{dm2}) and mixing matrix elements
of eq. (\ref{mns}).

There are no right-handed neutrinos in the SM. Neutrinos can have
a Majorana mass, but they are mass protected by electroweak gauge
invariance and the SM Higgs mechanism does not generate a Majorana
mass; a weak isotriplet Higgs boson would be required. However
physics beyond the SM could easily generate an effective
non-renormalisable interaction of the form
\begin{equation}
 \frac{1}{\Lambda}\phi_{WS}\phi_{WS}LL
\end{equation}
where $L$ is a SM lepton doublet and $\mu=\Lambda$ is the energy
scale where the SM breaks down. This interaction would generate a
neutrino Majorana mass term
\begin{equation}
 m_{\nu}= \frac{2v^2}{\Lambda}
 \label{effective}
\end{equation}
However we should note that, if there are some mass protecting
chiral charges beyond the SM, the neutrino mass could be
suppressed further and eq. (\ref{effective}) really provides an
order of magnitude upper limit to the neutrino mass. If we take
$\Lambda = \Lambda_{GUT} \sim 3 \times 10^{16}$, this upper limit
on the Majorana neutrino mass is $m_{\nu} \sim 0.002$ eV. This is
at least an order of magnitude less than the neutrino mass
required to explain atmospheric neutrino oscillations, see eq.
(\ref{dm2}). So there may well be new physics below the SUSY-GUT
scale responsible for neutrino masses and, more generally, for the
effective light neutrino Majorana mass matrix:
\begin{equation}
\overline{\nu_{Li}} (M_{\nu})_{ij} \nu_{Lj}^c
\end{equation}
Fermi-Dirac statistics means that $M_{\nu}$ must be symmetric.

If we assume that, as in many models, the charged lepton mass
matrix is quasi-diagonal then the dominant contributions to the
neutrino mixing matrix $U_{MNS}$ come from the Majorana neutrino
mass matrix $M_{\nu}$. In this case there are two leading order
forms \cite{altarelli,king} for $M_{\nu}$, according to whether
neutrino masses have a hierarchical or an inverse hierarchical
spectrum\footnote{We do not consider here the fine-tuned
possibility of quasi-degenerate neutrino masses.}. Hierarchical
models, with neutrino masses satisfying $|m_3| \gg |m_2|, |m_1|$,
have a leading order mass matrix of the type:
\begin{equation}
M_{\nu} \sim \pmatrix{0       & 0       & 0\cr
                      0       & 1       & 1 \cr
                      0       & 1       & 1\cr}\frac{m}{2}
\end{equation}
Inverted hierarchical models, with $|m_1| \approx |m_2| \gg
|m_3|$, have a leading order mass matrix of the form:
\begin{equation}
M_{\nu} \sim \pmatrix{0       & 1       & 1\cr
                      1       & 0       & 0 \cr
                      1       & 0       & 0\cr}\frac{m}{\sqrt{2}}
\end{equation}
where $m \simeq \sqrt{|\Delta m_{23}^2|} \simeq 0.05$ eV.

The best-known scenario for generating the effective light
neutrino mass matrix from beyond the SM physics is the see-saw
mechanism \cite{seesaw}. Here one introduces three right-handed
neutrinos, which are not mass protected by the electroweak
interactions. Hence they are expected to have a symmetric Majorana
mass matrix $M_R$, with right-handed neutrino mass eigenvalues
characteristic of some new physics energy scale---the see-saw
scale. In addition one expects a normal electroweak scale Dirac
neutrino mass matrix $M_N$, connecting the left-handed and
right-handed neutrinos, as in eq. (\ref{eq:ab2}) for the $SO(10)$
model considered in the previous section. An effective light
neutrino Majorana mass matrix is then naturally generated and
given by the see-saw formula:
\begin{equation}
M_{\nu} = M_N M_R^{-1} M_N^T
\end{equation}
In the above $SO(10)$ model, the see-saw formula generates
realistic neutrino masses with a right-handed neutrino see-saw
scale $\Lambda_R = 3 \times 10^{14}$ GeV and a large solar
neutrino mixing angle \cite{geer}. However the large atmospheric
neutrino mixing angle is generated from the charged lepton mass
matrix $M_E$ of eq. (\ref{eq:ab1}), which is not quasi-diagonal.
It is also possible to fit the neutrino masses and mixing angles
in the AGUT model \cite{fnt}, in which $M_E$ is quasi-diagonal.

\section{Conclusion}
\label{conclusion}

We have reviewed some approaches to the quark-lepton mass problem.
Of course we did not have time to cover the whole field and, for
example, did not discuss the interesting ideas from models with
extra dimensions \cite{arkani} for understanding fermion masses.
The hierarchical structure of the spectrum was emphasized and
interpreted as due to the existence of a mass protection
mechanism, controlled by approximately conserved chiral flavour
quantum numbers beyond the SM. Many flavour symmetry models have 
been constructed; however it must be admitted that none of them 
is totally convincing. The existence of {\em two} large neutrino 
mixing angles was not predicted and provides a challenge for 
model builders. So I leave you with the message that there is still 
much exciting research to be done---both theoretical and 
experimental---in the search for the physics of flavour underlying 
the Standard Model. 

\section*{Acknowledgements}
I would like to thank Michael Danilov and the Organising Committee
for their warm hospitality at this enjoyable and interesting School.

\end{document}